\documentclass[lettersize,conference]{IEEEtran}
\usepackage{amsmath,amsfonts}
\usepackage{algorithmic}
\usepackage{algorithm}
\usepackage{array}
\usepackage[caption=false,font=normalsize,labelfont=sf,textfont=sf]{subfig}
\usepackage{textcomp}
\usepackage{stfloats}
\usepackage{url}
\usepackage{verbatim}
\usepackage{graphicx}
\usepackage{cite}
\usepackage{bm}
\usepackage{xspace}
\usepackage{hyperref}

\newcommand{\eg}{e.g.,\xspace}

\begin{document}

\title{A Fast Locality Simulator for GEMM Design-Space Exploration on Multi-Chiplet GPUs}

\author{
    \IEEEauthorblockN{
        Euijun Chung,
        Hyesoon Kim
    }
    \IEEEauthorblockA{
        Georgia Institute of Technology
    }
    \IEEEauthorblockA{
        \emph{
        \href{mailto:euijun@gatech.edu}{euijun@gatech.edu},
        \href{mailto:hyesoon@cc.gatech.edu}{hyesoon@cc.gatech.edu}
        }
    }
}

\maketitle

\begin{abstract}
In multi-chiplet GPUs, memory accesses that cross the silicon interposer to a remote chiplet's high-bandwidth memory (HBM) incur extra latency and energy, making remote-traffic reduction crucial for efficiency.
For general matrix multiply (GEMM), the dominant operator in LLMs, inter-chiplet traffic depends strongly on design knobs such as per-operand memory layout, cooperative thread array (CTA) traversal order, and data placement. 
The optimal combination is difficult to identify analytically, as locality depends strongly on CTA traversal and its interaction with the L2 cache.

To this end, we present a fast, tile-level locality simulator that models data placement and CTA-to-chiplet mapping, CTA traversal, per-chiplet L2 caches, and local/remote HBM accesses.
This enables rapid evaluation of locality, performance, and energy efficiency under various GEMM configurations.
Using the simulator, we find that the best locality-aware configuration for each GEMM reduces remote traffic by up to 18.3$\times$ and improves energy efficiency by up to 17\% over 4\,KB-interleaved data with round-robin CTA-to-chiplet mapping.
Moreover, using the simulator output as feedback, an agentic AI adopts a 2D block-swizzle CTA traversal that improves mean energy efficiency by 15.2\% for Qwen and 6.9\% for Llama relative to the best 1D traversal under 4\,KB-interleaved data placement. 
Overall, our simulator enables fast exploration of the GEMM locality design space on multi-chiplet GPUs and is available at \url{https://github.com/gthparch/chiplet_locality_simulator}.
\end{abstract}

\section{Introduction and Motivation}

\begin{figure}[t]
    \centering
    \includegraphics[width=1.0\linewidth]{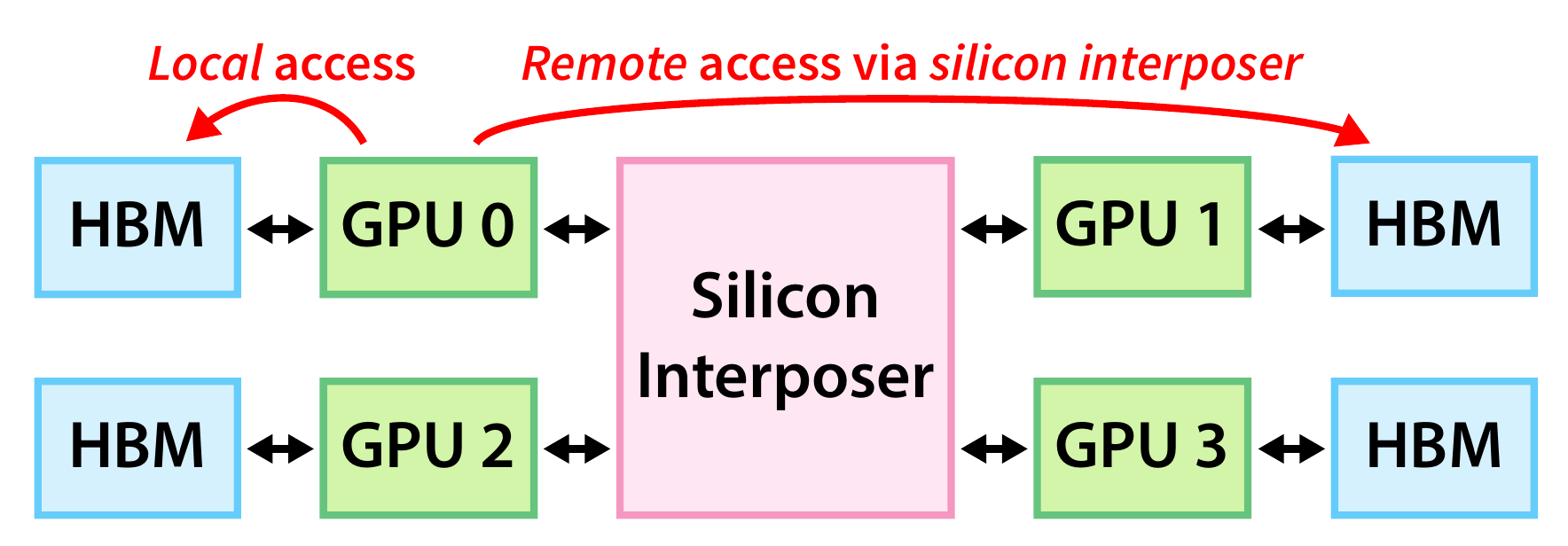}
    \vspace{-0.3in}
    \caption{A 4-chiplet GPU. Remote access across the interposer increases latency, energy consumption, and inter-chiplet bandwidth usage.}
    \label{fig:chiplet}
\end{figure}

Multi-chiplet GPUs scale compute and memory capacity beyond the reticle limit, but memory accesses within a package become either \emph{local} or \emph{remote} to each chiplet~\cite{arunkumar2017mcm, amd_cdna3_architecture_2025, nvidia_blackwell_architecture_2024}. 
As shown in Figure~\ref{fig:chiplet}, a memory request is served either by a near HBM (\emph{local}) or a far one across the interposer (\emph{remote}); remote accesses add latency and energy consumption, and consume shared inter-chiplet bandwidth~\cite{tee2025mall}. 
Improving \emph{chiplet locality}, the fraction of traffic served locally, is therefore central to chiplet-GPU efficiency~\cite{park2025leveraging, chowdhary2026fleet, soria2025delegato, seyyedaghaei2025memory, dalmia2024cpelide}.

\begin{figure}[t]
    \centering
    \includegraphics[width=1.0\linewidth]{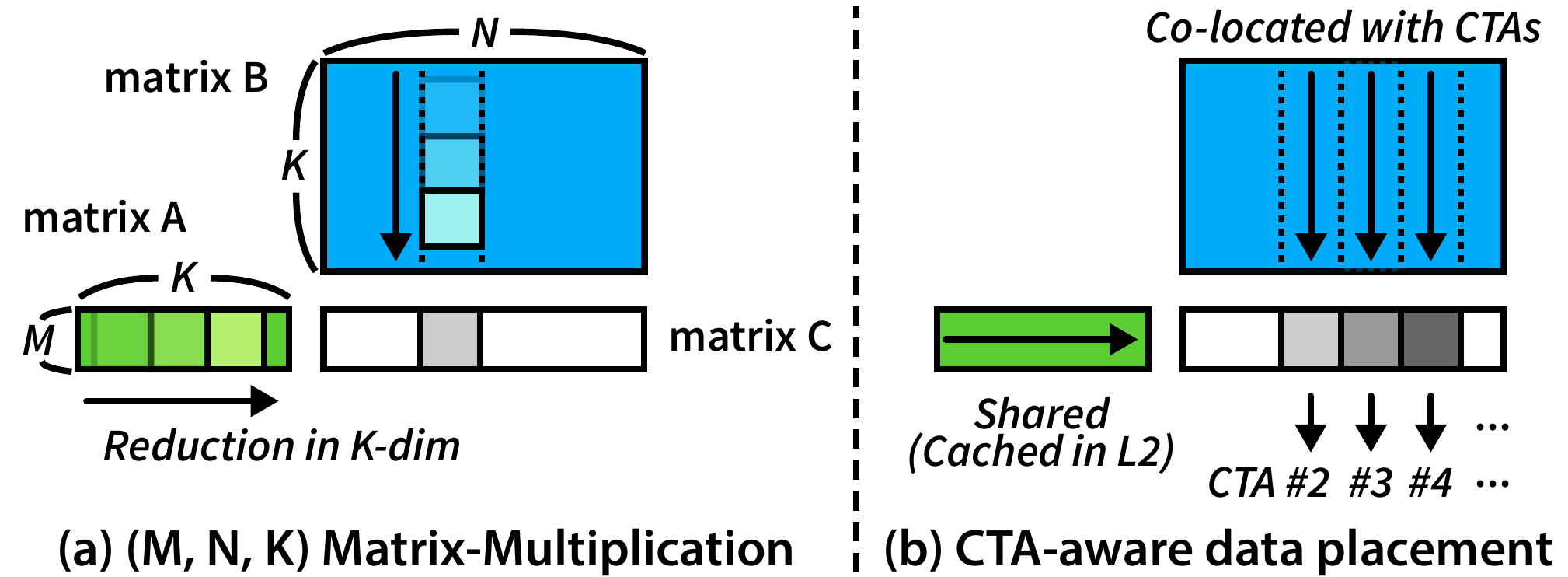}
    \vspace{-0.25in}
    \caption{(a) Tiled matrix-multiplication. (b) Co-locating an operand region with the CTAs that consume it, while caching the shared operand in L2, reduces remote HBM accesses.}
    \label{fig:gemm}
    \vspace{-0.1in}
\end{figure}

GEMM (General Matrix Multiply) operations dominate LLM training and inference, and their tiled execution exposes predictable CTA (cooperative thread array)-to-operand affinity (Figure~\ref{fig:gemm}). 
Locality-aware data placement and CTA-to-chiplet mapping exploit this affinity by co-locating each CTA with the data it consumes~\cite{kim2018coda, khairy2020locality, joo2026characterizing, feng2024barre, green2026nuna, zhang2026location, bolado2025rex}. 
However, the knobs that control locality span a large design space: per-operand row/column-major layout, CTA traversal order, and the placement of input and output data across chiplets. 
Although a compiler can predict accesses as local or remote from the layout and CTA mapping~\cite{khairy2020locality}, actual remote HBM traffic is determined after per-chiplet L2 filtering. 
CTA traversal and cache replacement therefore make the outcome difficult to predict with a simple static rule or closed-form analytical model, and the best configuration varies significantly across GEMM shapes~\cite{zhang2023sac, xu2025exploiting, mallya2025memplex}.
For instance, for a single representative GEMM in LLM training, remote HBM traffic ranges from roughly 0.26\,GB to 11.4\,GB across the data placement and CTA-to-chiplet assignment policies.

Finding the optimal configuration for these knobs is challenging on today's hardware and software. 
Vendor-tuned GEMM implementations are often proprietary and difficult to reproduce exactly, and hardware profilers expose aggregate cache and memory counters but do not attribute traffic to local versus remote chiplet HBM at the granularity needed to compare placements and traversals~\cite{liu2026dgna, ryu2026understanding}. 
Cycle-level simulators can in principle measure this~\cite{khairy2020accel, sun2019mgpusim, chung2026macsim}, but are impractically slow for the large GEMMs common in LLM workloads (\eg one Qwen3 GEMM has $(M, N, K) = (49{,}152,\, 16{,}384,\, 2{,}048)$)~\cite{yu2026patterns, chung2025swift}.
Moreover, their detailed kernel and hardware models make broad sweeps over locality configurations computationally expensive and cumbersome.

Our goal is instead to model the components that directly determine chiplet locality while abstracting instruction-level timing. 
GEMMs have a regular tiled execution in which each CTA computes one output tile, allowing the simulator to enumerate each tile's memory accesses without simulating individual arithmetic instructions. 
To this end, we developed a lightweight functional simulator for design-space exploration of GEMM on multi-chiplet GPUs.
Our simulator models data accesses through the per-chiplet L2 caches under the CTA schedule and data placement to determine local and remote HBM traffic, while the performance is estimated using a first-order roofline model and dynamic energy using per-operation and per-access energy costs.

\begin{figure*}[t]
    \centering
    \includegraphics[width=1.0\linewidth]{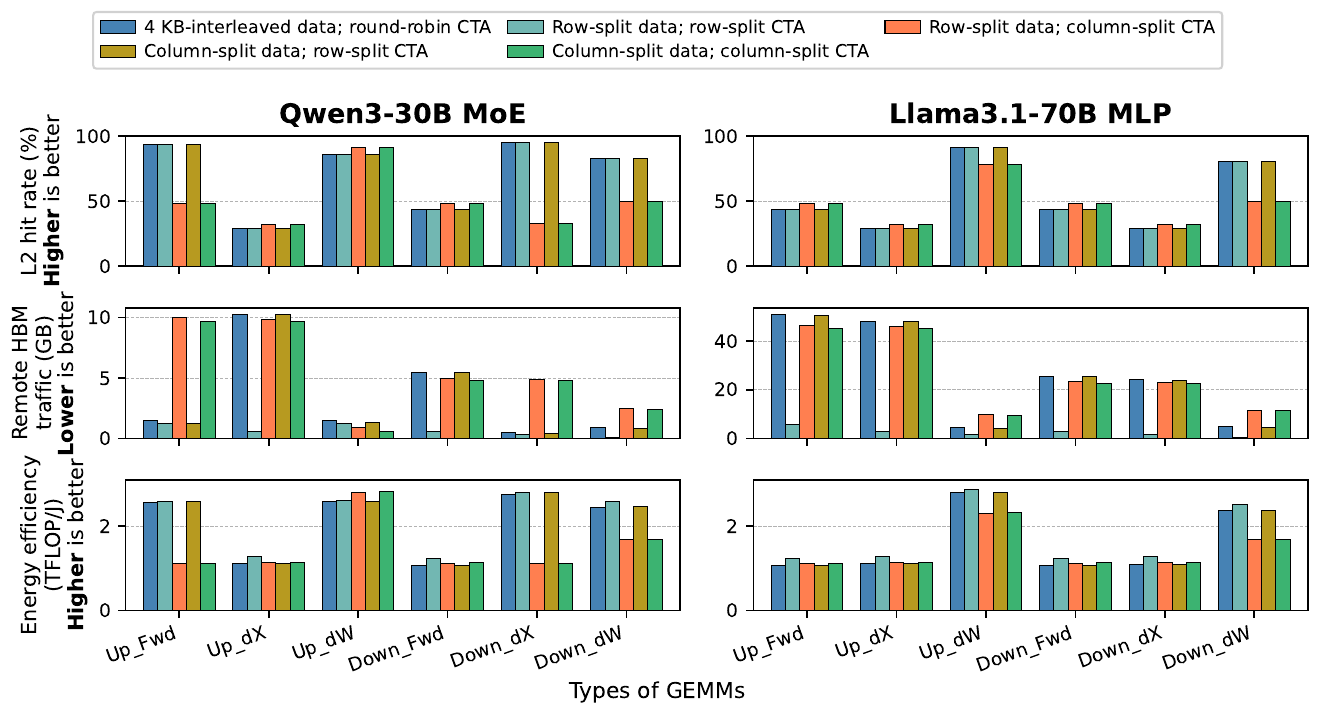}
    \vspace{-0.3in}
    \caption{    
    L2 hit rate, remote HBM traffic, and energy efficiency for 12 LLM GEMMs with 4K total tokens. 
    We compare four combinations of row-/column-split A/B placement and CTA-to-chiplet assignment against 4 KB-interleaved data with round-robin CTA assignment.
    Up/Down denote the gate-up and down projections; \texttt{Fwd}, \texttt{dX}, and \texttt{dW} denote the forward and two backward GEMMs.
    }
    \label{fig:design}
\end{figure*}

\section{GEMM Locality Simulator}

We model one MI300X-like package as four HBM locality domains, each with 76 compute units (CUs) and a private 8\,MB, 16-way L2 (128\,B lines). 
The simulator is \emph{functional}: each CTA computes one output tile and issues operand accesses determined by the GEMM shape, tile size, and operand layout, in the selected CTA traversal order.
Accesses are filtered through the per-chiplet L2, and every L2 miss and output write is classified as local or remote based on the data-placement policy. 
Each CTA holds its A/B operand tiles in shared memory, so per-CU occupancy is set by the shared-memory budget.
CTAs within a wave advance through the K dimension in lockstep, issuing their strided A/B reads through the chiplet's L2 at each K-step before any CTA proceeds.

The simulator knobs are (i) per-operand row/column-major layout; (ii) CTA traversal order for the output matrix, either 1D row- or column-major or a 2D block-swizzle; and (iii) data placement for the input (A/B) and output (C) matrices, either 4\,KB round-robin interleaving or locality-aware row-/column-split placement. 
Row-split placement divides each logical row across the four chiplets along its column dimension, whereas column-split placement divides each logical column along its row dimension.
For locality-aware placement, CTA-to-chiplet mapping follows the split of the output matrix. 
For 4\,KB-interleaved output data, CTAs are instead assigned round-robin across chiplets in the selected traversal's dispatch order.

The outputs of the simulator are L2 hit rate, HBM locality $=$ local$/$(local$+$remote), absolute inter-chiplet (remote) traffic, and roofline-based performance and dynamic-energy estimates. 
L2 hit rate is measured over A/B reads; C writes bypass L2, and remote HBM traffic counts remote A/B L2 misses and remote C writes.
The simulator models data movement rather than cycle-level timing. 
Performance is estimated with a roofline model with MI300X-class compute throughput and HBM bandwidth, and dynamic energy is derived from per-operation and per-access energy costs~\cite{o2017fine, rusu2023ucie, naffziger2021pioneering}.
Execution time is estimated as the maximum of compute time and HBM-transfer time, and energy efficiency is computed as total FLOPs divided by dynamic energy.
Dynamic energy includes per-FLOP compute energy, HBM-access energy, and additional interconnect energy for remote traffic.
We do not model the memory-side last-level cache (\eg AMD Infinity Cache), since it resides beyond the interposer crossing and therefore does not affect the local/remote classification of traffic, only the fraction that ultimately reaches DRAM.

\section{Results and Insights}

\begin{figure}[t]
    \centering
    \includegraphics[width=\linewidth]{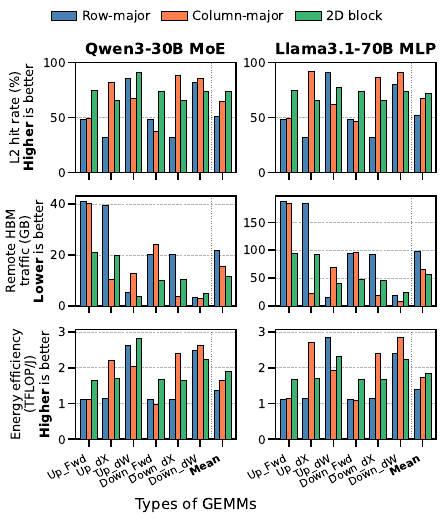}
    \vspace{-0.35in}
    \caption{    
    Impact of changing the GEMM kernel’s CTA traversal order with 16K total tokens. With 4 KB-interleaved data and a  round-robin CTA-to-chiplet mapping, we compare row-major, column-major, and 2D block-swizzle traversal using L2 hit rate, absolute remote HBM traffic, and energy efficiency.}
    \label{fig:traversal}
\end{figure}

\noindent \textbf{Workloads.} We evaluate representative GEMMs from Qwen3-30B-A3B (mixture-of-experts, MoE) and Llama3.1-70B (dense), in both the forward and backward pass~\cite{yang2025qwen3, grattafiori2024llama}. 
Figure~\ref{fig:design} reports the FFN gate-up and down projections.
We also evaluated the fused QKV (Query, Key, Value) and O (Output) projections and observed similar qualitative sensitivity to data placement and CTA mapping; we omit them for brevity.
Each FFN projection appears as a forward (\texttt{Fwd}) and two backward (\texttt{dX},\texttt{dW}) GEMMs.
We use a tile size of $(t_m, t_n, t_k) = (128, 128, 32)$ and \texttt{BF16} operands. 

\noindent \textbf{Data \& CTA placement policies.} In Figure~\ref{fig:design}, \emph{row-/column-split data} refers to the placement of A and B, while \emph{row-/column-split CTA} refers to the CTA-to-chiplet mapping induced by C placement. All configurations use row-major CTA traversal. We use 4 KB-interleaved data with round-robin CTA-to-chiplet mapping as a conventional baseline~\cite{amd_cdna3_architecture_2025}.

\noindent \textbf{Performance and dynamic energy estimation.}
We assume effective memory-access energies of 3.5 pJ/bit for local and 4.9 pJ/bit for remote accesses~\cite{o2017fine, rusu2023ucie, naffziger2021pioneering}.
We further assume 0.3 pJ/FLOP for BF16 computation, 1307.4 TFLOP/s compute throughput, 5.3 TB/s HBM bandwidth, and 8.1 TB/s interconnect bandwidth.

\noindent \textbf{L2 cache behavior.}
We assume that each chiplet's private L2 caches both local and remote A/B reads, while streaming C writes bypass L2 and are included directly in HBM traffic.

\noindent \textbf{Results.} Among the five mappings in Figure~\ref{fig:design}, the choice of data placement and CTA-to-chiplet mapping changes remote HBM traffic by up to 43.1$\times$ for a single 4K-token GEMM. 
The best combination differs across GEMMs, so no single fixed mapping is optimal.
Changing CTA traversal adds a further lever (Figure~\ref{fig:traversal}). 
Relative to 4\,KB-interleaved data with round-robin CTA-to-chiplet mapping, the evaluated locality-aware mappings reduce remote traffic by up to 18.3$\times$ and improve energy efficiency by up to 17\%.
The simulator also shows that mismatching the data split and CTA-to-chiplet mapping can increase remote HBM traffic by up to 17.0$\times$.
Each full-size 16K-token GEMM configuration simulates in 9--44\,s (mean: 24\,s) on a single CPU core, making full design-space sweeps practical. 

\noindent \textbf{Agentic AI use-case.} We then allowed an AI agent (Claude Code with Opus 4.7) to modify and run our simulator, tasked with reducing a GEMM's remote traffic by tuning the simulator knobs. 
Exploring only the CTA traversal knob under fixed hardware with 4\,KB-interleaved data and round-robin CTA-to-chiplet mapping, the agent adopted a 2D block-swizzle order, a standard L2-locality optimization in dense-GEMM kernel libraries such as CUTLASS and Triton~\cite{nvidia2026cutlass, choudhary2025optimizing}.
Figure~\ref{fig:traversal} reports the evaluation results. 
Compared with the best fixed 1D traversal, 2D block-swizzle traversal increases mean L2 hit rate by 9.4 and 4.4 percentage points, reduces mean remote HBM traffic by 25.9\% and 13.4\%, and improves mean energy efficiency by 15.2\% and 6.9\% for Qwen and Llama, respectively. 
The per-GEMM effect is mixed: block traversal reduces remote traffic for 5 of the 12 GEMMs, but it provides the best aggregate L2 hit rate, remote traffic, and energy efficiency for both models.
CTA traversal order is a first-order, shape-dependent control for inter-chiplet traffic that AI and automated kernel generators should account for. 
Moreover, because the simulator exposes both hardware and software knobs in one fast model, it makes joint hardware/software optimization easier to explore.

\section{Impact and Limitations}

Our simulator enables fast, AI-driven design-space exploration of chiplet-GPU locality for GEMMs.
Its main limitation is that it is functional: performance is estimated using a first-order roofline model, while energy is derived from analytical per-operation/access costs rather than cycle-accurate measurements.
The execution model also assumes wave-synchronous (lockstep) CTA progress, which can slightly perturb the modeled L2 hit rate. 
In our evaluated configurations, the A/B footprint issued by one wave at each K-step is smaller than the per-chiplet L2, so we expect this effect on locality to be small.
While our evaluation selects the best policy for each GEMM independently, because tensors may be reused across GEMMs with different preferred policies, end-to-end policy selection is left to future work.
Overall, the simulator turns chiplet-GPU GEMM locality into a fast, transparent design space that engineers and AI agents alike can explore in tens of seconds.

\clearpage

\bibliographystyle{IEEEtran}
\bibliography{refs}

\end{document}